\documentclass[journal,article,submit,moreauthors,pdftex]{article}

\usepackage[english]{babel}
\usepackage{xcolor}
\usepackage[utf8]{inputenc}
\usepackage[T1]{fontenc}
\usepackage{subcaption}
\usepackage{hyperref}
\usepackage{nameref}
\usepackage[normalem]{ulem}
\usepackage{amssymb}
\usepackage[inline]{enumitem}
\usepackage{placeins}
\usepackage{authblk}

\usepackage[letterpaper,top=2cm,bottom=2cm,left=3cm,right=3cm,marginparwidth=1.75cm]{geometry}

\usepackage{amsmath}
\usepackage{graphicx}

\title{Infrastructures connecting people.
A mechanistic model for terrestrial transportation networks.}

\author[1,2]{Luce Prignano}
\author[3]{Lluc Font-Pomarol}
\author[1,2]{Ignacio Morer}
\author[1,4]{Sergi Lozano}

\affil[1]{Universitat de Barcelona Institute of Complex Systems (UBICS) Universitat de Barcelona, Barcelona, Spain}
\affil[2]{Departament de Fisica de la Mat\`eria Condensada, Universitat de Barcelona, Barcelona, 08028, Spain}

\affil[3]{Departament d'Enginyeria Química, Universitat Rovira i Virgili, Tarragona, Spain}
\affil[4]{Departament d'Hist\`oria Econ\`omica, Institucions, Pol\'itica i Economia Mundial, Universitat de Barcelona, Barcelona, 08028,Spain}

\newcommand{\ie}{\emph{i.e.}, }
\newcommand{\eg}{\emph{e.g.}, }

\newcommand{\avg}[1]{\langle #1 \rangle}
\newcommand{\Eloc}{E_{\text{loc}}}
\newcommand{\Eglob}{E_{\text{glob}}}
\newcommand{\Pmin}{P_{\text{min}}}
\DeclareMathOperator{\sign}{sign}

\begin{document}
\maketitle

\begin{abstract}
The structure and evolution of Terrestrial Transportation Infrastructures (TTIs) are shaped by both socio-political and geographical factors, hence encoding crucial information about how resources and power are distributed through a territory. Therefore, analysing pathway, railway or road networks allows us to gain a better understanding of the political and social organization of the communities that created and maintained them.
Network science can provide extremely useful tools to address quantitatively this issue. Here, focusing on passengers transport, we propose a methodology for mapping a TTI into a formal network object able to capture both the spatial distribution of the population and the connections provided by the considered mean of transport. Secondly, we present a simple mechanistic model that implements a wide spectrum of decision-making mechanisms which could have driven the creation of links (connections). Thus, by adjusting few parameters, for any empirical system, it is  possible to generate a synthetic counterpart such that their differences are minimized. 
By means of such inverse engineering approach, we are able to shed some light on the processes and forces that moulded transportation infrastructures into their current configuration, without having to rely on any additional information besides the topology of the network and the distribution of the population.
An illustrative example is also provided to showcase the applications of the proposed methodology and discuss how our conclusions fit with previously acquired knowledge (and literature) on the topic.
\end{abstract}

\section{Introduction}

Historically, transportation infrastructures have had huge impact over the development of the territories. They allow the movement of people and goods, affecting the long-term capability to cope with changing socio-economic scenarios. Among them, terrestrial transportation infrastructures (hereafter referred as TTIs) differ from other transportation systems mainly in that their connections are physical.

This implies that, unlike in the case of air, maritime, or river transportation, whose growth concentrates the costs on localized structures --such as airports or harbours-- to enable the hosting of more connections, most of the costs of TTIs are associated to the construction of connections themselves --roads, railways, etc.-- while a much smaller fraction of the total budget is usually for toll booths or stations. 

This particular feature opens the door to the exploitation of TTIs as a fundamental source of information about the societies that created and maintained them. 

Indeed, only in the case of terrestrial infrastructures, the decision of connecting two previously disconnected places is a crucial, not easily reversible one. The balance between the cost of building a new connection -- which typically increases with the distance -- and the provided benefit to the system affects the development of TTIs, which can hence be regarded as the result of multiple interests and competing constraints. 

Their structure and evolution are influenced, on the one hand, by the changing needs that they are supposed to satisfy, and on the other hand, by how resources and power are distributed through a territory \cite{herranz2017, FPM2021}. 

In this sense, each TTI as a whole is the emerging output of a complex process involving many interacting actors and different spatial and temporal scales.

In the context of archaeological research, the importance of TTIs for the understanding of the political and social organization of the societies that created and maintained them was initially assessed in relation to the Roman Empire, and more recently in a number of studies in the New World and in Pre-Roman Europe (see, for instance, \cite{chevallier_1976, taylor_1979, trombold_1991a, jenkins_2001, smith_2005, Groenhuijzen_2016}). 

In previous works, building on this literature, we took a further step and tried to infer aspects of the political organization of a region from the quantitative analysis of TTI. 

We proposed a methodology to assess the relationship among political entities from the structure of transportation networks in some proto-historical case-studies \cite{prignano-jas-2019, fulminante_etal_2017, fulminante_etal_2021}.

Here, focusing on passengers transport (PTTIs), we present a generalised version of such a methodology, extending its applicability to current-time case-studies. In particular, this generalised approach allows to address questions like to what extend an infrastructure was conceived as an independent system rather than as an auxiliary network that complements another; or whether a particular PTTI was shaped according to the demographic spatial  distribution.

The article is organised as follows. The Second Section summarizes our baseline methodology.

Afterwards, we discuss how to adapt both ingredients to the characteristics of current-time scenarios. In Section \ref{sec:ingredient_1}, we focus on how to process empirical data about PTTIs that are very different in nature from that of 'archaeological PTTIs'. For instance, present case studies are much richer in details whose relevance needs to be assessed and that could be treated in multiple ways. 

In Section \ref{sec:ingredient_2}, we discuss what kind of hypotheses are suitable to be taken into consideration in modern and contemporary scenarios, connecting them with relevant research questions and integrating the knowledge available about the administrative and institutional organization of the systems under study. Such a reflection is translated to specific modifications to the original approach.

In Section \ref{sec:results_discussion}, we discuss the results of applying our methodology to an illustrative case-study: the Catalan railway system, a regional PTTI with several decades of history that coexists with a road system. Special attention is paid to the assessment of the performance of the models and to how to draw conclusions properly from the comparison of empirical and synthetic data. Finally, the article closes with some concluding remarks. Our preliminary results show how this approach has good potential, calling for further research.

\section{Baseline methodology}\label{sec:baseline}

In previous works, we tackled the issue of quantitatively 
inferring aspects of the political organization of a region
from the structure of transportation networks \cite{prignano-jas-2019, fulminante_etal_2017, fulminante_etal_2021}. 

Our goal was to identify the nature of the interplay between different human communities, going beyond tasks usually addressed by archaeological quantitative analysis, such as establishing the existence of a certain degree of regional organization.

To this aim, we addressed the analysis of the decision-making processes prioritizing some paths over others by envisioning a methodology consisting of two fundamental ingredients:

\begin{enumerate*}[label=(\roman*)]
\item a procedure for extracting relevant quantitative data from road maps; 
\item formal models implementing alternative mechanisms for generating synthetic TTIs to be compared against the empirical ones. 
\end{enumerate*}

The underlying idea is that some models reproduce relevant structural features of the empirical TTI with higher accuracy than others. In other words, they provide a higher quality explanation of the empirical evidence and, therefore, we assume them to be more likely to resemble the actual mechanisms of regional organization \cite{prignano-jas-2019}.

We adopted network science as a natural framework to address the interplay between connectivity and functionality of TTIs. Indeed, network science provides us both with tools to identify and measure structural characteristics of empirical TTIs and with a conceptual framework for formal model building \cite{boccaletti_2006,Newman2003,Fontoura2011}.

The task of translating road maps into networks is not straightforward and can be performed in many alternative, not equivalent ways. Since we were studying inter-settlement interactions, we needed our nodes to represent the human communities connected through the regional TTI. Then, as the simplest possible option, we established a bidirectional link between any two sites that were directly connected by a terrestrial route, with no other settlement in between. To include the geographical factor in a simple way, we represented sites as geo-localized nodes and assigned weights to the links according to the geodesic distance between the nodes they connected.

We designed a minimalist set up in which each node, at each step, expressed a preference concerning the new link to be established, according to a variably well informed assessment of costs and benefits. Then they could either compete against each other or reach an agreement about which connection was going to be built next. 

It is worth stressing that our goal was not to understand why in a certain region there were more or less settlements, or more or less roads. On the contrary,  we were addressing the question of why and how the settlements that existed in the region built those roads instead of others. Consequently, the models took the set of settlements with their corresponding geographic locations and the amount of available resources -- here quantified as the total link length $L_{tot}$ -- as inputs, not as parameters to be fitted.

The process ended when the total length of the connections added was equal to the total link length of the corresponding empirical network. Consequently, any synthetic graph generated by a network model replicated the following characteristics of the corresponding empirical network:
\begin{enumerate*}[label=(\arabic*)]
    \item the total number of nodes $N$,
    \item its geographic density $\delta=L_{tot}/\sum_{i=1}^N\sum_{j=i+1}^N d_{ij}$, and 
    \item the average node strength $\avg {s} = \frac{1}{N} \sum_{i=1}^{N}\sum_{j\in V} l_{ij}$, where $V$ is the set of neighbors of $i$ and $l_{ij}$ is the length of the link between $i$ and its neighbor $j$.
\end{enumerate*}

Any other metric is, in principle, suitable to be used for comparing different synthetically generated graphs against their empirical counterpart.

\subsection{The Equitable Efficiency Model (EE Model).}
\label{sec:EEM}
Out of all the models that we devised, one presented a higher explanatory power when applied to our proto-historical case studies (\ie it generated the synthetic networks more similar to the corresponding empirical ones).
It equipped nodes with global information about their connectivity, but also with the ability to make coordinated decisions. More concretely, it assumed that each settlement knew the (weighted) length of each one of the existing path joining it with any other settlement. Then links were prioritized globally according to their normalized distance $R$ that was calculated as follows:

\begin{equation} \label{eq:detour}
R_i(j) = R_j(i) = R_{ij} = \frac{d_{ij}}{L_{ij}},
\end{equation}

where $d_{ij}$ is the geodesic distance between node $i$ and node $j$, and $L_{ij}$ is the length of the shortest existing path between them. The normalized distance $R$ can also be seen as the inverse of what is called the route factor or detour index \cite{barthelemy_2011}. If $i$ and $j$ are disconnected, \ie belong to different connected component, this means that there exist no finite length path between them. Hence 

\begin{equation} \label{eq:02}
R_{ij}^{[d]}=\lim_{L_{ij}\to \infty} \frac{d_{ij}}{L_{ij}} = 0
\end{equation}

and the comparison between disconnected node pairs is performed by determining

\begin{equation} \label{eq:03}
\sign \left( R_{ij}^{[d]} - R_{lm}^{[d]}\right) =
\sign \left(\lim_{L^{[d]}\to\infty}\frac{d_{ij}-d_{lm}}{L^{[d]}} \right) = \sign (d_{ij}-d_{lm})
\end{equation}
which is obtained assuming that the path length $L^{[d]}$ is the same for all the disconnected nodes.
 
The function $R_i(j)$ balances costs and benefits, prioritizing those links that shorten long paths (large $L_{ij}$) while wasting little resources (short $d_{ij}$).

More concretely, the EE model follows a three-step procedure:

\begin{enumerate}
    \item For each node $i$, all the $R_i(j)$ values are calculated. 
    \item Each node $i$ proposes the creation of a link between itself and a node $j^*$ such that the $R_i(j^*)$ was the minimum value among all the $R_i(j)$ (local interest expressed by node $i$).
    \item All the proposals are ranked according to their $R$ value and a link is created between the pair corresponding to the global minimum (coordinated decision-making).
\end{enumerate}

Step 1, 2, and 3 are repeated until the summed lengths of all created links reaches that of the empirical system\footnote{Such a procedure is equivalent to simply building, at each step, the link to the minimum of the $R_{ij}$ matrix. Nonetheless, the metaphor of the individual priorities that have to be sorted is useful for devising meaningful generalization of the present baseline model.}.

\subsection{EE model with preferential attachment}
\label{sec:EEM-pa}
The EE model considered all settlements to be on the same ground and the links to be built were selected among the individual preferences according to a fair criterion. In order to explore a slightly different scenario, where preferences of nodes with more and longer links were entitled to a higher priority level, we devised a variant of the EE model. Mathematically, such a bias is obtained by weighting the ratio $R$ with a (negative) power of the strength (or weighted degree, \ie the total length of its adjacent links) of the proposing node, thus introducing preferential attachment among the nodes with the greatest strength. The trade-off between the two ingredients in determining the priority of each link is tuned by the exponent a of such power. Hence, the new value of the biased ratio $R'$  for a connection between node $i$ and $j$ proposed by node $i$ is:
\begin{equation}\label{eq:04}
R'_{ij}=R_{ij} s_i^{-a}= \frac{d_{ij}}{{L}_{ij}}s_i^{-a} 
\end{equation}
where $s_i$ is the strength of node $i$. Therefore, when $a$ is equal to zero, we recover the EE Model.

It is worth noting that in all the scenarios considered, including the last one, we assumed that all node-settlements were intrinsically equally important. When prioritising new links to be established, the node-settlements based their choice on geographical (distances) and topological (already existing links) information, but on node-settlement attributes (such as power, richness or attractiveness).

\section{From maps to accessibility networks}\label{sec:ingredient_1}
In the case-studies considered, all of them from the Iron Age, the region under study was provided of a single PTTI that embedded the footprint of the relationships between the human communities living in the area. When human societies started building roads \cite{earle_1991,lay_1992}, they created, in each territory, a network made of a combination of artificially adapted natural paths and manufactured ways that served for displacements at all the scales --from local to supra-regional-- and for all means of transport (pedestrian, by wheeled vehicles, with animals). 
Then, each settlement had to be connected to others by means of this single PTTI, or it would be otherwise isolated.
This in not the case for later scenarios. Already in the Roman Empire, the most important cities were connected through primary roads, while less important towns and villages were reachable thanks to a dense net of secondary roads and less manufactured pathways (see, for instance, \cite{carreras-de-soto-2013}). Such a difference plays a crucial role both in the way the empirical network is constructed and in the modelling approach.

Considering their high construction costs, some PTTIs, such as railway systems or highways, are not designed to directly reach each single town and village in a territory. Instead, many human settlements benefit indirectly from the presence of a station or exit nearby.

Since one of our objectives is to capture the effect of population distribution over the territory on PTTIs structural patterns, the option to discard some groups of inhabitants based on whether their residence place is reached by the infrastructure under study or not must be ruled out.

We had therefore to disregard some of the most usual network representations, such as the so-called space-of-stations for railway systems, in which nodes are stations and links represent physical connections \cite{kurant_and_thiran_2006}.

As an alternative, we sought for a method to integrate information about both population distribution and PTTI's connectivity in a single empirically-based graph (\ie a PTTI accessibility network). Specifically, we developed a procedure to assign fractions of the population to each train station, highway exit, etc. (hereafter, referred as station for the sake of simplicity) by 'merging' neighbouring localities, while redefining both their coordinates and connectivity.

Our merging procedure grasps information about population distribution from the most fundamental units (\ie in general and hereafter, the municipalities) and their geographical positions. Initially, each municipality (with or without direct train connection) is represented as a node. The process takes into account population sizes and distance between municipalities to, iteratively, joining two nodes into one. In that way, low populated nodes are likely to be merged with closer, larger ones, in a location somewhere in the middle of both. Along the process, newly merged nodes preserve any connection existing in the previous step.

Formally speaking, the process is represented by an algorithm that gets as input information about the initial set of nodes $i$ (geographic coordinates and population $P_i$) and the links among them, and executes the following steps:

\begin{enumerate}
\item  It creates a list with all possible pairs of nodes $i,j$ computing its inter-distances $d_{ij}$. The list is ranked in ascendant order according to the distances.
\item For the first element in the list, it calculates $\Gamma_{ij} = \Pmin * d_{ij}$ , where $\Pmin = \min{(P_i,P_j)}$. This quantity expresses the balance between the importance of the smallest node and the distance that separates it from the largest one.
\item If the condition $\Gamma_{ij} < \Gamma^*$ is fulfilled, both nodes merge into one.

The new node will have a population $P^\prime = P_i+P_j$ and a new position (Lat', Lon') at the center of mass of $i$ and $j$ according to their former populations:
\begin{equation}
 Lon^\prime=\frac{P_i Lon_i + P_j Lon_j}{P_i+P_j} \quad \text{,} \quad
 Lat^\prime=\frac{P_i Lat_i + P_j Lat_j}{P_i+P_j}
 \label{eq:new_pos}
\end{equation}

\item If there existed a link between the two merged nodes, the link disappears. If there were links adjacent to $i$ or $j$, they are rewired correspondingly to the new node. Notice that those links would also change their length since the position of the merged node is now somewhere between former $i$ and $j$ positions.
\item The four previous steps are repeated until no more pairs of nodes are able to merge.
\end{enumerate}

The parameter $\Gamma^*$ fixes how lowly populated a node has to be and how close to another one more populated to merge them. It acts as a restrictive merging condition, being more permissive to nodes' merging as it increases its value. By means of the parameter $\Gamma^*$, we are considering the finite capacity of a station, allowing for a higher node density in highly populated areas, a necessary feature of almost any PTTI. 

\begin{figure}
\centering
\begin{subfigure}{0.8\textwidth}
  \centering
  \includegraphics[width=0.8\linewidth]{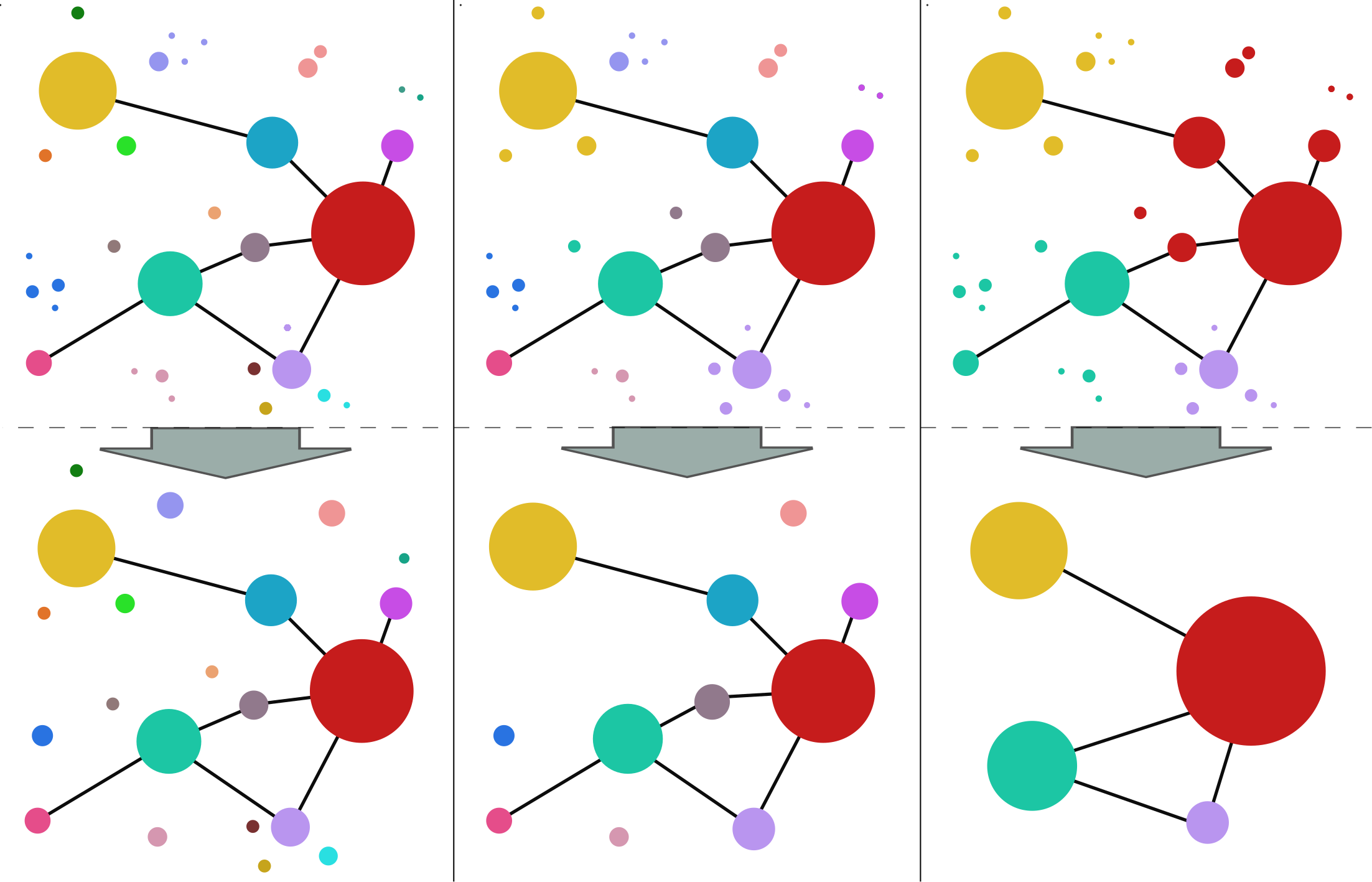}
 \label{fig:merging_output}
\end{subfigure}%
\caption{Schematic example of the merging procedure for the same initial network and a small (left), intermediate (centre), and large (right) value of $\Gamma^*$. Node size proportional to the population, colors according to merging results.}
\end{figure}

This method presents several advantages. It is independent on the territory and it is able to capture potential heterogeneity across a region (\eg rural and urban). More importantly, this method is able to define the scale of the system that suits our goals. In this sense, it is more flexible and less arbitrary than allocating nodes based, for instance, on administrative divisions of the territory (\eg provinces, counties...) and allows for an easier comparison between case studies.

Although different solutions are possible, in most of the cases it is sensible to set the value of $\Gamma^*$ so that in the output the number of links matches, or slightly exceeds, the number of nodes. In principle, this condition allows to obtain a single connected component. Hence, isolates - if any - can be interpreted as representative of the population and geographical position of less favoured communities in terms of accessibility to the PTTI under study. Selecting a lower $\Gamma^*$ would make the number of nodes to be greater than the number of links, thus forcing some nodes to stay disconnected and making this feature less meaningful. On the contrary, a considerably higher $\Gamma^*$ might hide the existence of purely connected sub-regions by forcing further node merging.

\section{Generalized modelling approach}\label{sec:ingredient_2}
The main difference between proto-historical scenarios and more recent case studies is the relevance of the context preexisting the construction of the infrastructure under study. The first steps of the algorithms of our baseline models consist in building links between isolated nodes that are completely equal, except for their geographical coordinates. The underlying hypotheses is that differences in power and importance did not preexisted the creation of the very first transportation infrastructure, before which there were just a number of (almost) disconnected settlements. However, if we want to model a PTTI that is not the first ever built communication network in the region, this supposition does not hold.
In particular, we must take into account (1) the coexistence of more PTTIs, which implies that there exist alternative ways to reach each location, and (2) the uneven level of agency throughout the different communities in the considered territory. Hence, nodes in the initial state of the models are neither equal, nor disconnected. 
Here we address such issues keeping the algorithms as well as the input data as simple as possible.

\subsection{Generalised Equitable Efficiency Model}\label{sec:gEEM}

Taking Eqs. \ref{eq:04} and \ref{eq:02} as a starting point, we introduce two modifications and define a new, generalized version of the EE model:
The \emph{generalized Equitable Efficiency Model} (hereafter referred as gEEM).
We are interested in including the preferential attachment mechanism, since it proved to be useful when handling unbalanced settlements in terms of power\cite{fulminante_etal_2017}. However, instead of measuring the relative relevance across nodes using the node weighted degree (as in Sec.\,\ref{sec:EEM-pa}), gEEM uses an attribute of nodes that is now assumed to be a feature previous to the construction of the PTTI. Since our interest in this paper is focused on population distribution, here we are taking nodes' population. Notice, however, that this modelling approach could be applied using other attributes (such as nodes' wealth, for instance) depending on the requirements of each particular application. Thus, a new definition for the normalized distance is proposed:
\begin{equation} \label{eq:R_pop}
R_{ij}^\prime=\frac{d_{ij}}{L_{ij}} (p_i)^{-a} \quad \text{,} \qquad p_i=\frac{P_i}{P_{min}}  
\end{equation}
where $P_i$ is the population (or any other desired attribute) of node $i$ and $P_{min}$ is the smallest population of all nodes.

On the other hand, we redefined the limit taken to evaluate the shortest path between two existing nodes, which the previous models took as $L_{ij}\rightarrow \infty$. We propose to make it finite defining it as:

\begin{equation}
L_{ij}^{[d]}=e^{-m}L_{CG} 
\end{equation}

where $i$, $j$ are nodes in different connected components and $L_{CG}$ is the total link length of a complete graph\footnote{A graph is said to be complete when it exists a link between each pair of nodes, creating a fully connected structure.} built on the same set of geolocalized nodes. The parameter $m$ controls the merging mechanism, making it less restrictive for larger values. By redefining this limit, our purpose is to reproduce the overall effect that other infrastructures have in shaping the railway network. Considering that between two disconnected nodes there is a finite existing path, we make them reachable. Thus, this trait allows to consider them not so primordial for the network and to open the possibility to construct other links beforehand (between already connected nodes, for instance).

\subsection{Network characterisation by structural metrics}\label{sec:metrics}

In order to better characterise the empirical network resulting from the merging process and compare it with synthetic counterparts later on, we calculated several network properties that provide information about both the structure of the network and its influence on communication dynamics:

\begin{itemize}

\item \emph{Average link length}: It is a useful metric for a basic characterization of the links in a weighted spatial network:
\begin{equation}
	\langle l_e \rangle=\frac{1}{N}\sum_{i=1}^{M} (l_e)_i  
  \end{equation}
where $M$ is the number of edges.  

Moreover, in order to get information about link length variability, we computed the standard deviation of this metric  $\sigma_{l_e}$

\item \emph{Standard deviation of node strength}: The average value of the node strength $\langle s \rangle$ is, by construction, the same for the empirical network as for any synthetically generated counterpart, but its standard deviation $\sigma_{s}$ can be used as an indicator of how balanced the node connectivity is.
\end{itemize}

On the other hand, we found reasonable to assess the efficiency of the empirical network in terms of its main functionality, namely the interchange of goods, information and people. Several ways to evaluate such property have been proposed \cite{latora_2001,vragovic_2005}. From the different definitions in this literature, we took one proposed specifically for weighted networks in \cite{vragovic_2005}. This definition compares the existing shortest path between two nodes, $L_{ij}$, with the geodesic distance that separates them, $d_{ij}$, assuming that the information flows better when the first value approximates the second. For a single connection, this definition coincides with our normalized distance $R$ and, as previously mentioned, can be seen as the inverse of what is called the route factor or detour index \cite{barthelemy_2011}. According to this approach, we calculated the global and local efficiencies of the network as follows:
\begin{itemize}
\item \emph{Global efficiency}: For each pair of nodes $i$ and $j$, it compares the path connecting them through the network ($L_{ij}$) with the ideal case corresponding to the straight line ($d_{ij}$). Then, the global efficiency is obtained by averaging this ratio over all pairs of nodes.
  \begin{equation}
  \Eglob=\frac{1}{N(N-1)} \sum_{i \neq j}^{N} \frac{d_{ij}}{L_{ij}}
  \end{equation}

\item \emph{Local Efficiency}: For each node $i$, it evaluates how well information is exchanged between its first neighbours when $i$ is missing. In other words, it assesses the robustness of the network and how it is able to deal with failures (for instance, when the communication through a specific node is not possible).
  \begin{equation}
  \Eloc=\frac{1}{N}\sum_{i=0}^{N} \frac{1}{k_i(k_i-1)} \sum_{j \neq k \in \Gamma_i}^{N} \frac{d_{jk}}{L_{jk/i}}
  \end{equation}
 
 Where $j$, $k$ belong to the subgraph of first neighbours of $i$, $\Gamma_i$, and $L_{jk/i}$ is the shortest path connecting $j$ and $k$ when the node $i$ is removed.
 
\end{itemize}

\section{Application of our generalised methodology to a present-day case study}\label{sec:results_discussion}

This section illustrates the applicability of our generalised methodology (including the merging procedure to obtain the empirical network and gEMM as the model generating synthetic counterparts) to present-day scenarios.

\subsection{Empirical accessibility network from node merging}
\label{sec:emp_acc_net}

The specific territory used in this study to illustrate the applicability of our generalised methodology was Catalonia, a Mediterranean region in northeast Spain. This region is densely populated and presents heavy unbalance between rural and urban areas, with more than 70\% of its population living in cities. It also has a relevant industrial sector and diverse and developed transportation infrastructures. Among them, the railway network. Such infrastructure can be divided into high-speed railroads and regional services. The former is actually part of a larger-scale structure, which has been framed according to a country-wide perspective (\ie the Spanish high speed rail service \textit{AVE}). Therefore, here we consider only the regional service, which is, composed of two structures managed by two different operators (\textit{ADIF} and \textit{FGC}) owned by the Spanish and Catalan regional governments, respectively.

\begin{figure}[ht]
\centering
\includegraphics[width=0.45\linewidth]{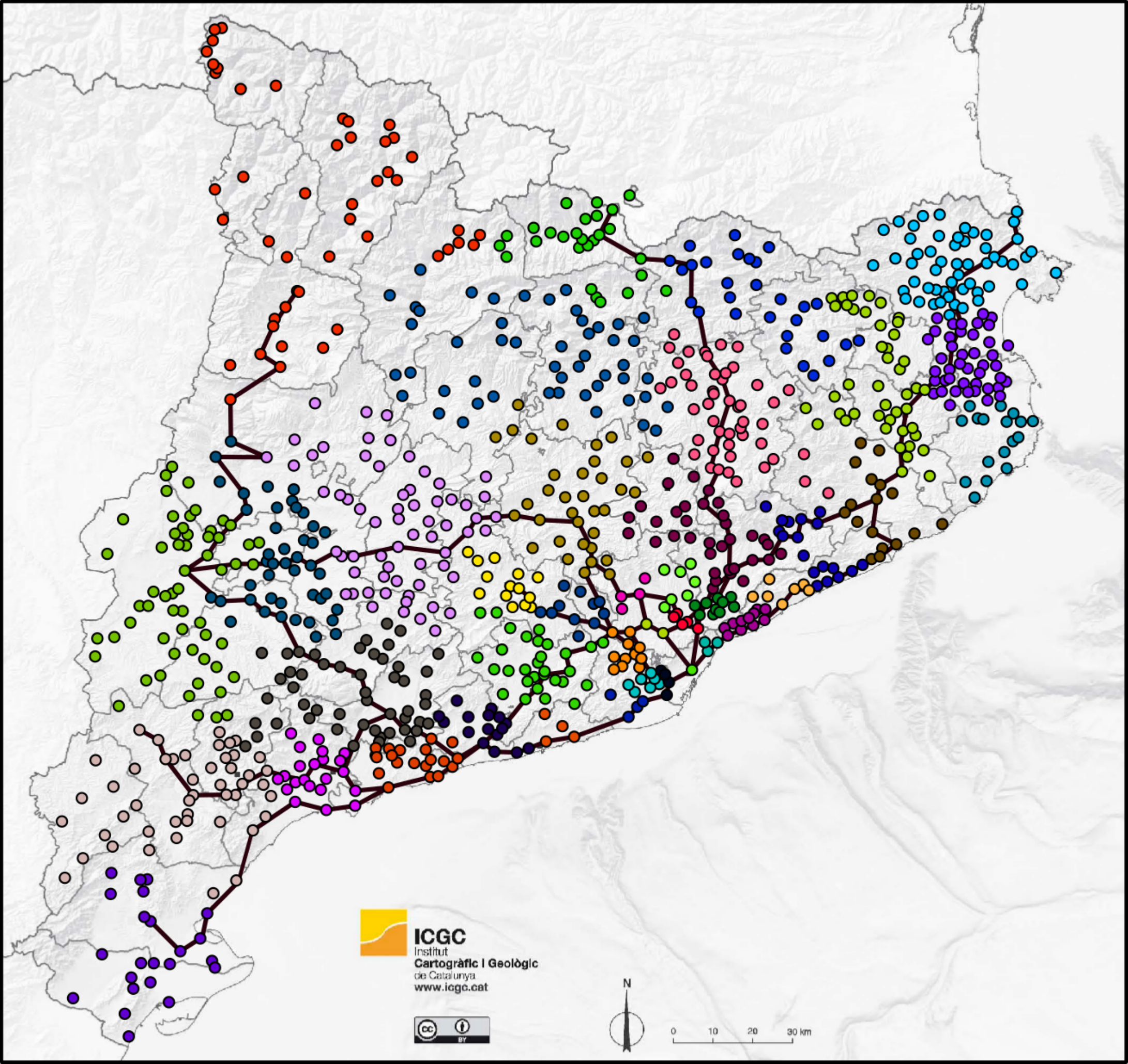}
\includegraphics[width=0.45\linewidth]{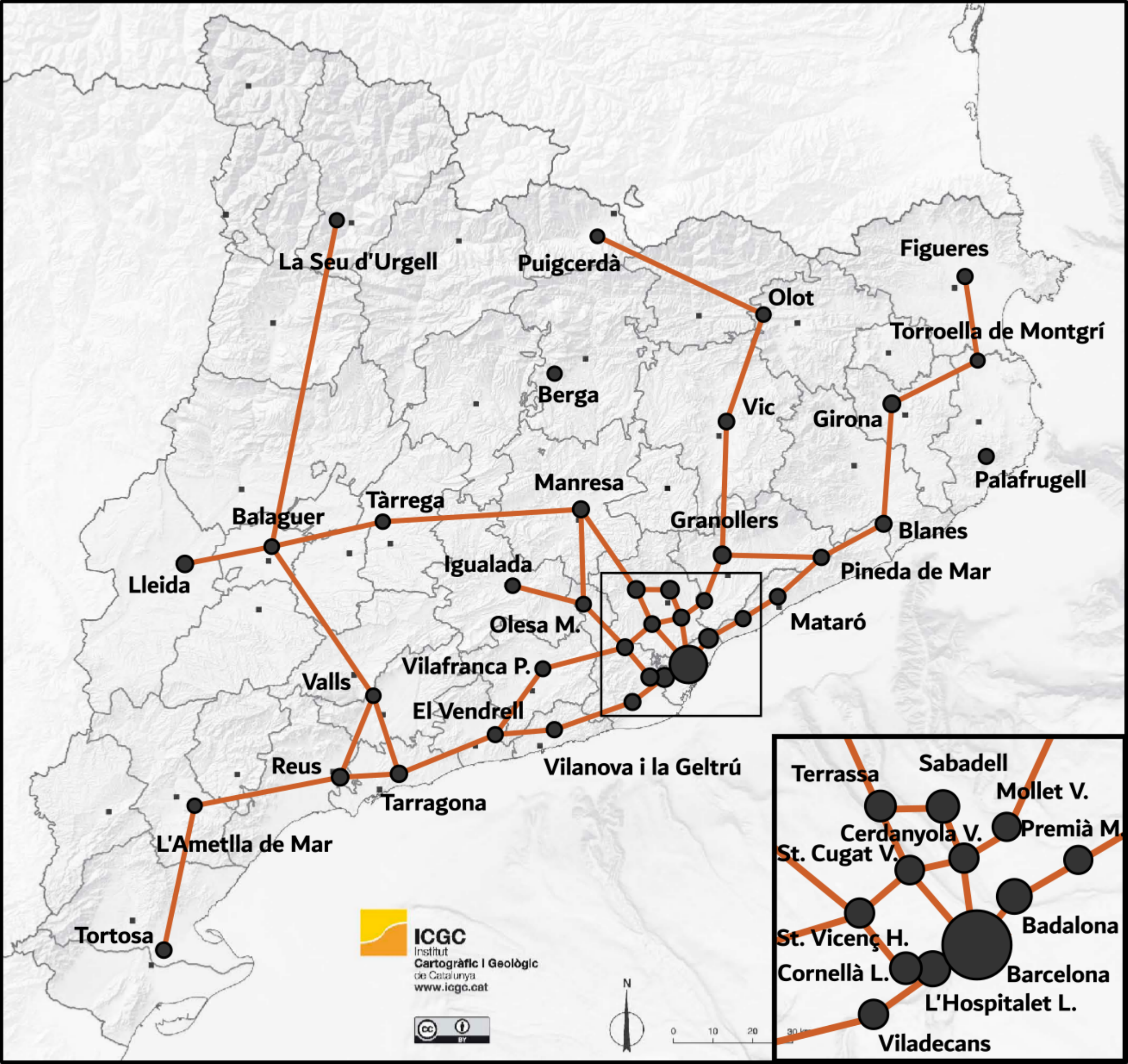}
\caption{Left: Municipalities ($N=945$) and train connections ($M=205$) before the merging procedure. The initial total link length is $L_{tot}^0=1208.51\, (km)$ and the total length of the corresponding MST is $L_{MST}^0=3737.26\, (km)$. Colours according to merging results. Right: Empirical network after merging procedure for $\Gamma^*= 0.82$. Labels assigned according to the most populated municipality. A node size is proportional to the total population of the municipalities inside it.}
\label{fig:empirical}
\end{figure}

In Figure \ref{fig:empirical} (left), we see Catalonia's regional railway networks, together with the municipalities served by them. As already pointed out in Section \ref{sec:ingredient_1}, many localities in Catalonia (and their corresponding share of the population) are not directly served by the railway networks. This justifies obtaining a train accessibility network as our empirical network instead of a physical railway one (\ie the space-of-stations representation). 

In this particular case, we chose to set the threshold parameter $\Gamma^*$ so that it is potentially possible for the resulting total link length to connect all the nodes in a single connected component. Starting from an initial state in which nodes outnumber links $945$ to $205$, the merging process progressively reduces the number of nodes, with some sporadic link losses as a collateral consequence. Meanwhile, the total link length of a hypothetical minimum spanning tree built on the same set of geolocated nodes also decreases, and does it faster than the total link length. Depending on the value of $\Gamma$, the resulting final network has different number of nodes $N(\Gamma)$ and links $M(\Gamma)$, and different total link length $L_{tot}(\Gamma)$ and minimum spanning tree length $L_{MST}(\Gamma)$. In general, the larger the value of $\Gamma$, the smaller the ratios $N(\Gamma)/M(\Gamma)$ and $L_{MST}(\Gamma)/L_{tot}(\Gamma)$, until they both reach a plateau. Once the condition $L_{MST}(\Gamma)/L_{tot}(\Gamma)<1$ is satisfied, the EE model always build a connected network, and actually if the ratio is equal to $1$ it produces the MST. However, in order to allow the present generalized version of the model to generate connected networks different from the MST, we chose a value $\Gamma^*$ such that the additional condition $M(\Gamma^*)/N(\Gamma^*)<1$ (\ie the number of links in the resulting empirical network exceeds the number of nodes) is also satisfied.

This is just a heuristic criterion, but has the clear advantage of not forcing the graph to have isolates while permitting their existence. In this way, the "surviving isolates" in the empirical network represent a meaningful trait of the system under study, not just a mere consequences of a wrong scale choice. We can thus apply our methodology to investigate the reason behind network features such as isolated notes, multiple connected components, and cycles (closed loops).
\begin{figure}[ht]
\centering
\includegraphics[width=0.6\linewidth]{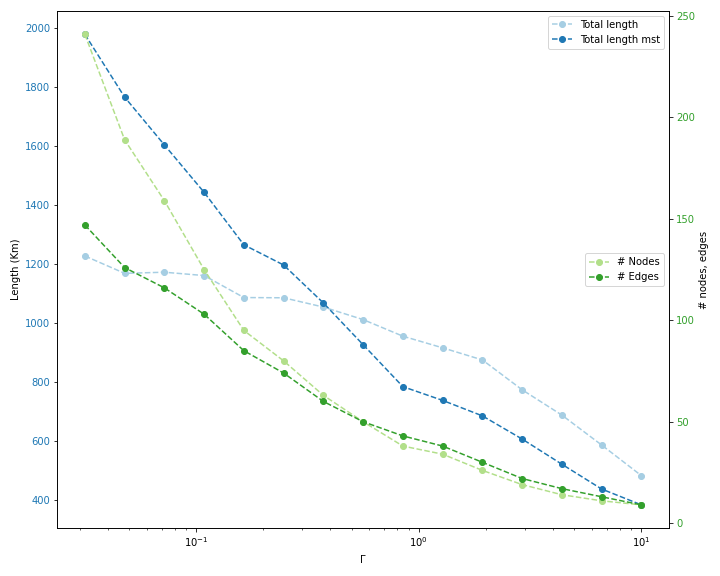}
\caption{The number of nodes $N$ and edges $M$ in the empirical network (right axis) and the corresponding total link length $L_{tot}$, together with the length of the minimum spanning tree $L_{MST}$ built on the same set of geolocalized nodes (left axis), are plotted as functions of the parameter $\Gamma$.}
\label{fig:length_nodes_edges}
\end{figure}

The empirical network is showed in Figure \ref{fig:empirical}, before (left) and after (right) applying the merging procedure setting $\Gamma^*=0.82$. 

\FloatBarrier

\subsection{Synthetic networks generated by the gEMM}
\label{sec:synt_acc_net}

Given a node layout $\mathcal{L}$, \ie set of geolocalized nodes, and a total link length $L_{tot}$ such that $L_{tot} > L_{MST}(\mathcal{L})$, by varying the two parameter $a$ and $m$, the gEMM is able to produce a great variety of network topologies, ranging from star-like graphs to disconnected cliques (all-to-all sub-graphs) and almost regular lattices, which may include a variable amount of isolates and connected components of different sizes.

To asses how some basic structural properties of the model's outputs depend on $m$ and $a$, we applied the gEMM to the node layout and total link length of the empirical network obtained in Sec.\,\ref{sec:emp_acc_net} varying both parameters. 

Since the number of different network topologies that can be constructed on a given node layout for a fixed total link length is finite, the parameter plane $m-a$ can be ideally partitioned into regions corresponding to an identical, unique output of the model. However, such an exhaustive exploration is beyond the scope of the present study. At a more coarse-grained scale, we observed that whenever $m \lessapprox 4.5$, this parameter does not affect the output of the model, that is, the output for any $(m,a)$ is the same as for $(4.5,a)$ for any $a$. Something analogous happens for the region $m \gtrapprox 10$ where the output for any $(m,a)$ is the same as for $(10,a)$ for any $a$. Similarly, for $a\gtrapprox 1$, the model does not produce any topological novelty and any network created for $(m,a)$, with $a>1$, is also generated for at least another pair $(m',a')$, where $m'<m$ and $a'<1<a$.

\begin{figure}
    \begin{subfigure}[h]{0.33\textwidth}
        \centering
        \includegraphics[width=\linewidth]{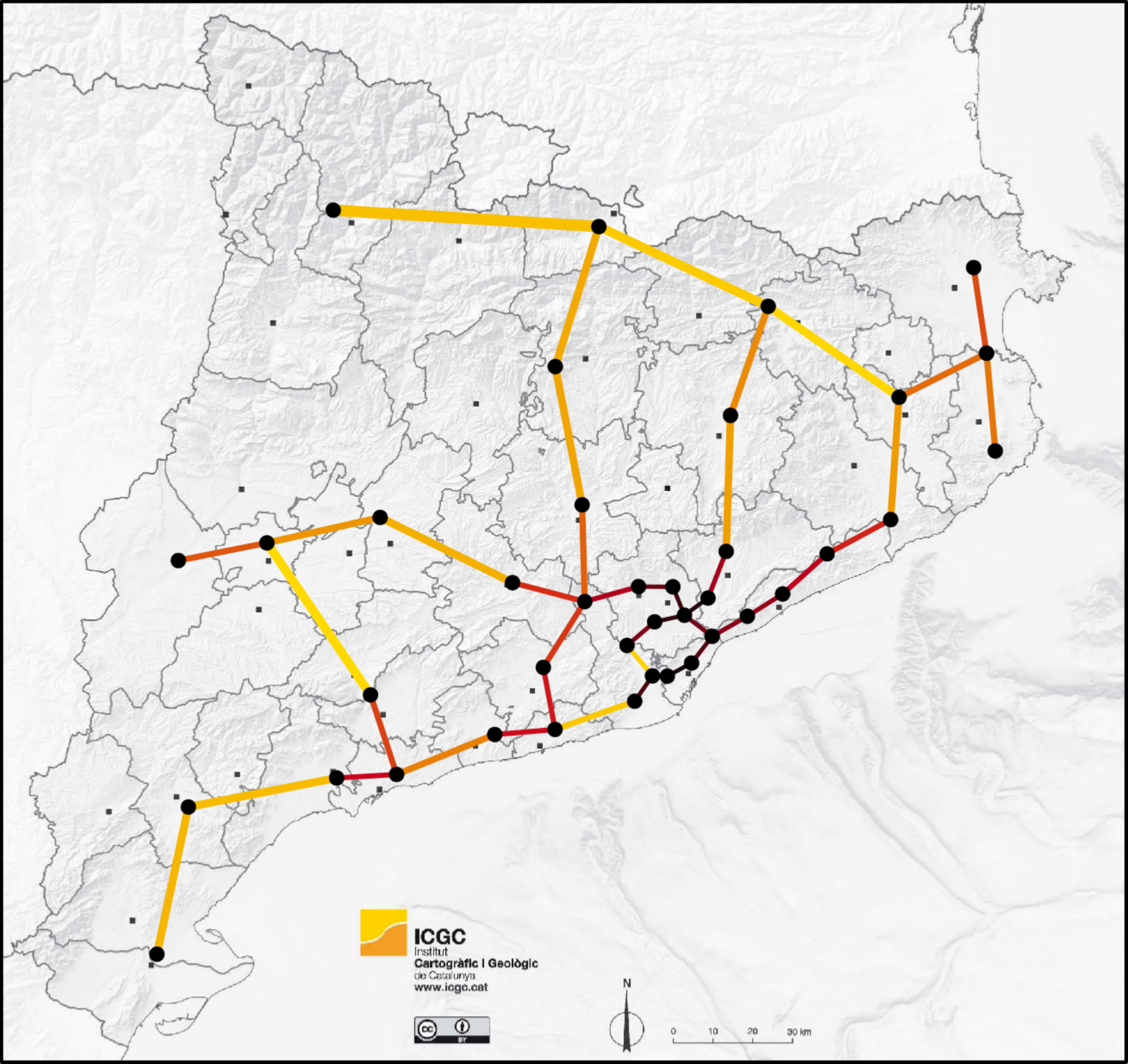}
    	\caption{$m=4.0$, $a=0.0$. (EEM model)}
    	\label{fig:model3}
    \end{subfigure}
    \begin{subfigure}[h]{0.33\textwidth}
        \centering
        \includegraphics[width=\linewidth]{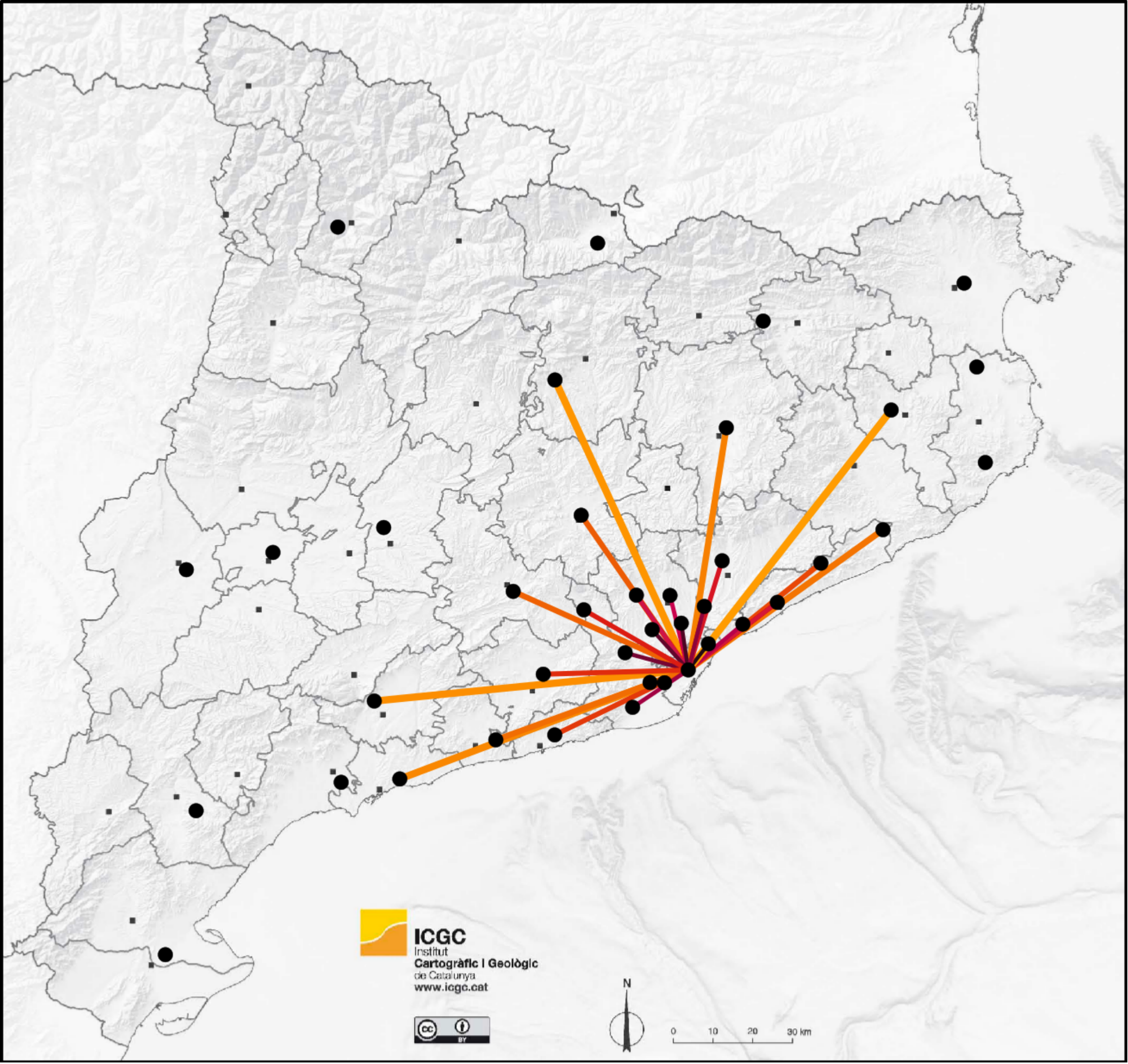}
    	\caption{$m=4.0$, $a=1.0$. }
    	\label{fig:sy2}
    \end{subfigure}
    \begin{subfigure}[h]{0.33\textwidth}
        \centering
        \includegraphics[width=\linewidth]{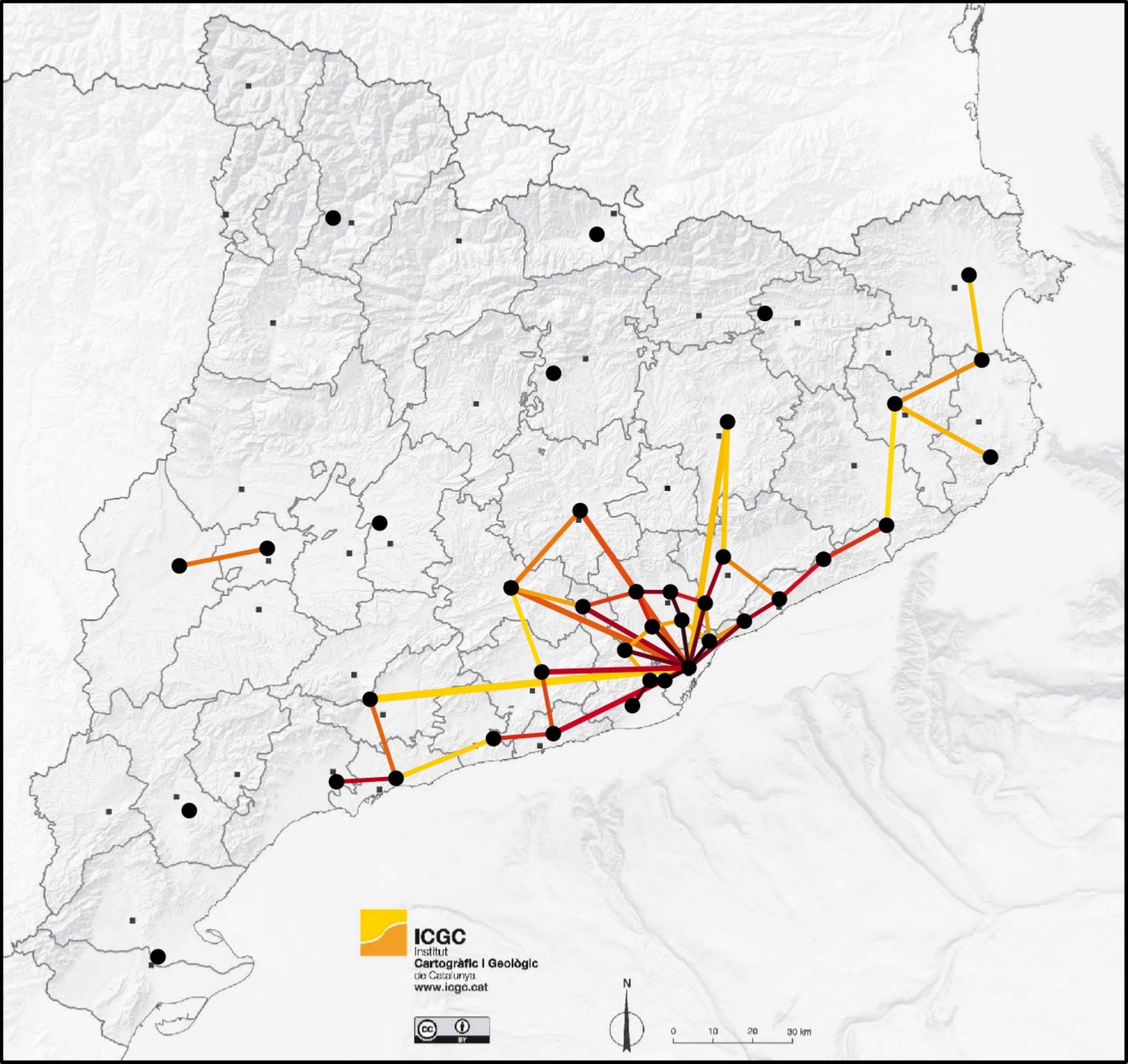}
    	\caption{$m=6.7$, $a=0.41$}
    	\label{fig:sy3}
    \end{subfigure}
    
    \vspace{0.8cm}
    \begin{subfigure}[h]{0.33\textwidth}
        \centering
        \includegraphics[width=\linewidth]{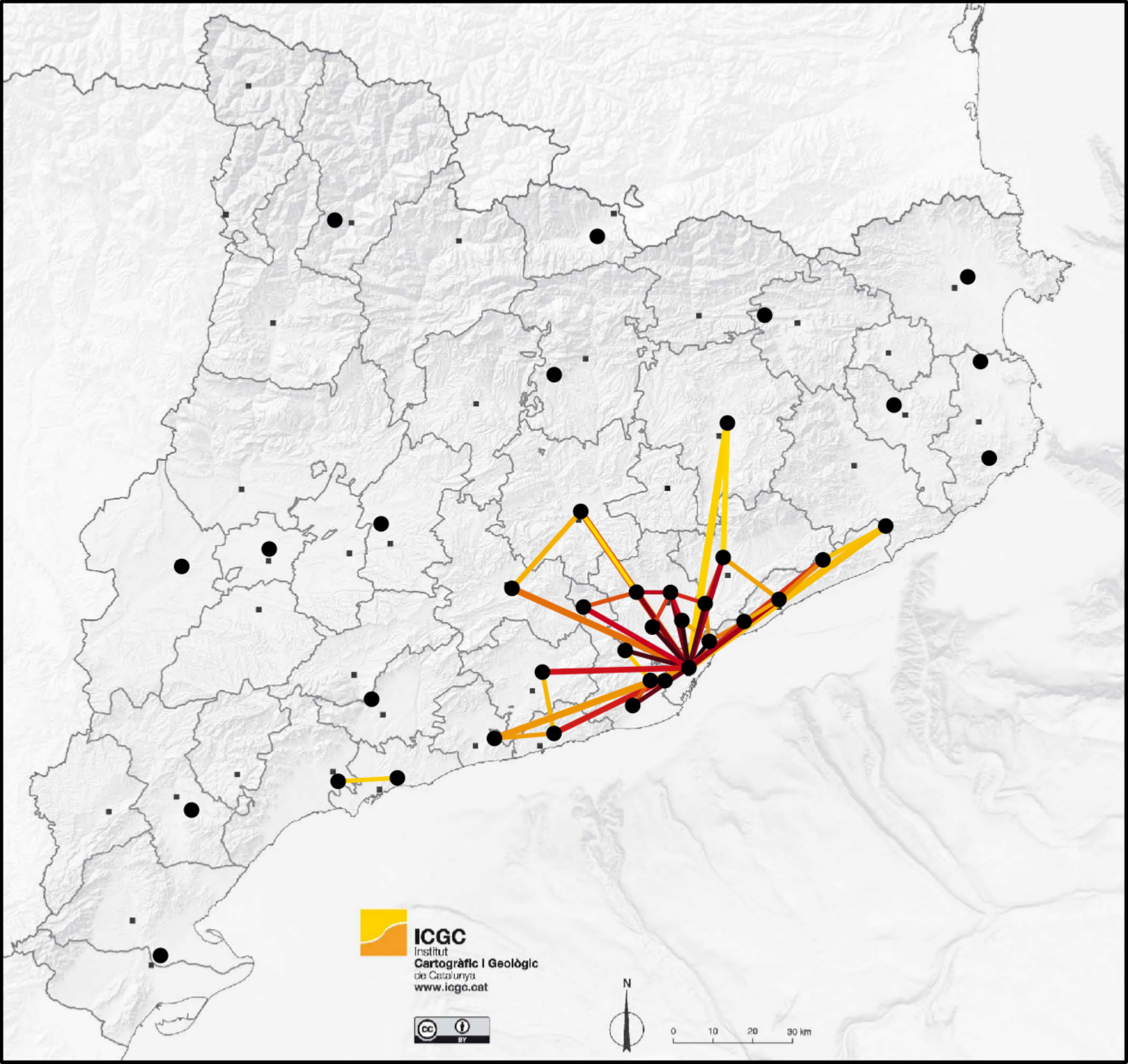}
    	\caption{$m=7.2$, $a=0.71$}
    	\label{fig:sy4}
    \end{subfigure}
    \begin{subfigure}[h]{0.33\textwidth}
        \centering
        \includegraphics[width=\linewidth]{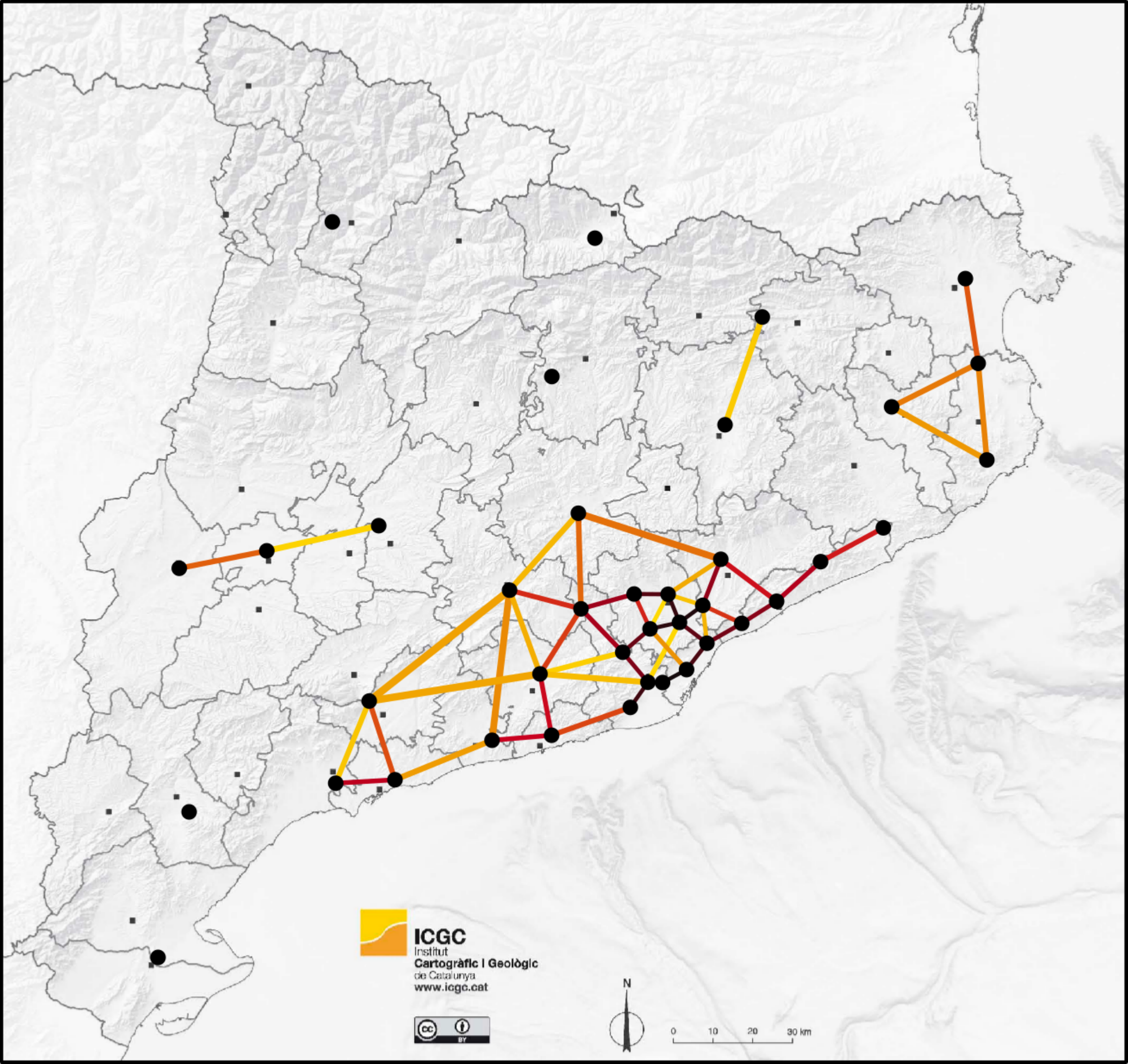}
    	\caption{$m=7.2$, $a=0$}
    	\label{fig:sy5}
    \end{subfigure}
    \hfill
    \begin{subfigure}[h]{0.33\textwidth}
        \centering
        \includegraphics[width=\linewidth]{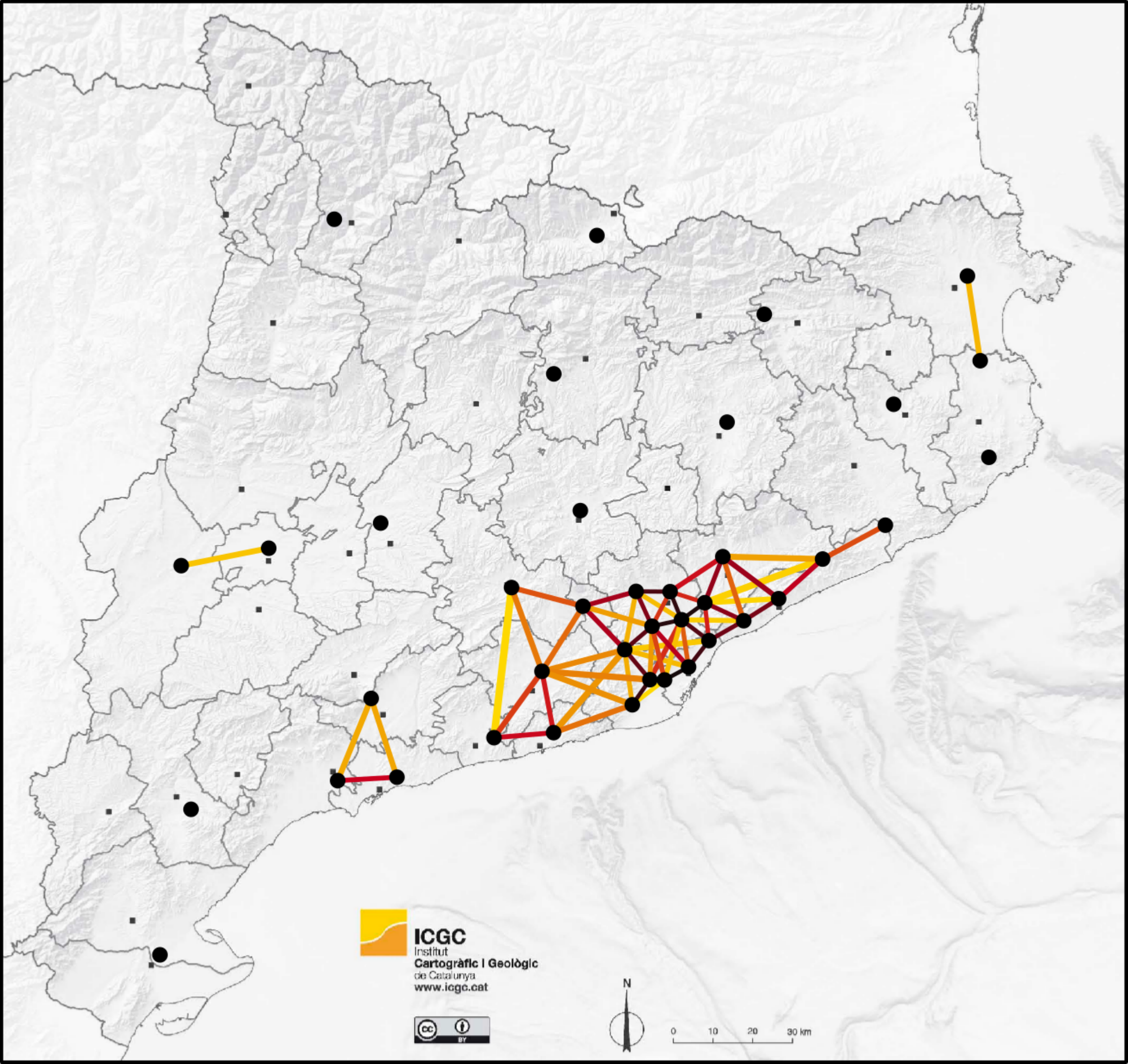}
    	\caption{$m=7.7$, $a=0.$}
    	\label{fig:sy6}
    \end{subfigure}
    \caption[Artificial networks examples.]
    {Artificial networks examples. \par \small Artificial networks for different pairs of values for the parameters are represented. Edges' color illustrates the order of construction of the links by the algorithm. Darker colors stand for earlier edges while brighter stand for later ones.}
    \label{fig:s_examples_1}
\end{figure}

In particular, for $m\lessapprox 4.5$ and $a=0$, the output of the gEEM is the same as that of the original EE model (Fig.\,\ref{fig:model3}). The first connections to be created belong to the MST, afterwards the algorithm adds few shortcuts until the total link length equals that of the empirical network.
When the value of any of the two parameters is increased, some of the links belonging to the MST start to be rewired and some nodes may be left disconnected from the largest component of the resulting network. While the principal effect of increasing the value of $m$ (Figs.\,\ref{fig:sy2},\ref{fig:sy3},\ref{fig:sy4},\ref{fig:sy5},\ref{fig:sy6}) is the appearance of closed triangles, larger values of $a$ (Figs.\,\ref{fig:sy2},\ref{fig:sy3},\ref{fig:sy4}) force the links to be redirected towards most populated nodes.

\FloatBarrier
\subsection{Comparing synthetic and empirical networks}
\label{sec:comparison}

\begingroup
\setlength{\tabcolsep}{10pt}
\renewcommand{\arraystretch}{1.5}
\begin{table}[h!]
\centering
 \begin{tabular}{||c c|| c c c c c c c ||} 
 \hline
 $m$ & $a$ & $\langle C \rangle$ & $\langle l_e \rangle$ & $\sigma_{l_e}$ & $\sigma_s$ & $I$ & $\Eglob$ & $\Eloc$ \\ [0.5ex] 
 \hline\hline
  $\leq4.0$ & 0 & 0 & 22.1 & 13.5 & 30.5 & 0 & \textbf{0.800} & 0.192 \\
  $\leq4.0$ & 1.0 & 0 & \textbf{36.9} & \textbf{24.6} & \textbf{153.1} & 12 & 0.358 & 0 \\
  6.7 & 0.41 & 0.120 & 29.8 & 20.8 & 125.2 & 7 & 0.336 & 0.123 \\
  7.2 & 0.71 & \textbf{0.431} & 21.8 & 15.2 & 103.0 & 14 & 0.431 & 0.281 \\
  7.2 & 0 & 0.208 & 18.2 & 9.63 & 37.4 & 5 & 0.208 & \textbf{0.389} \\
  7.7 & 0 & 0.360 & 15.9 & 6.67 & 42.6 & 11 & 0.360 & 0.285 \\
  $\geq 10.0$ & 0 & 0.295 & 16.0 & 6.45 & 75.3 & \textbf{27} & 0.295 & 0.089 \\
  $\geq 10.0$ & 1.0 & 0.299 & 16.7 & 7.25 & 77.6 & \textbf{27} & 0.299 & 0.086 \\

  \hline\hline
 \multicolumn{2}{||c||}{\textit{Empirical}} & 0.039 & 21.7 & 14.9 & 33.1 & 2 & 0.732 & 0.245\\ [1ex] 
 \hline
 \end{tabular}
 \caption[Network metrics values.]
    {Network metrics values. \par \small For different representative pairs of values for $m$, $a$, the corresponding metrics are shown, along with those corresponding to the empirical merging network. All metrics have no dimension except for $\langle l_e \rangle$, $\sigma_{l_e}$ and $\sigma_s$, which are expressed in $Km$. In bold, the highest value for each metric in the table.}
 \label{tab:properties}
\end{table}
\endgroup

\begin{figure}
    \begin{subfigure}[h]{0.33\textwidth}
        \centering
        \includegraphics[width=\linewidth]{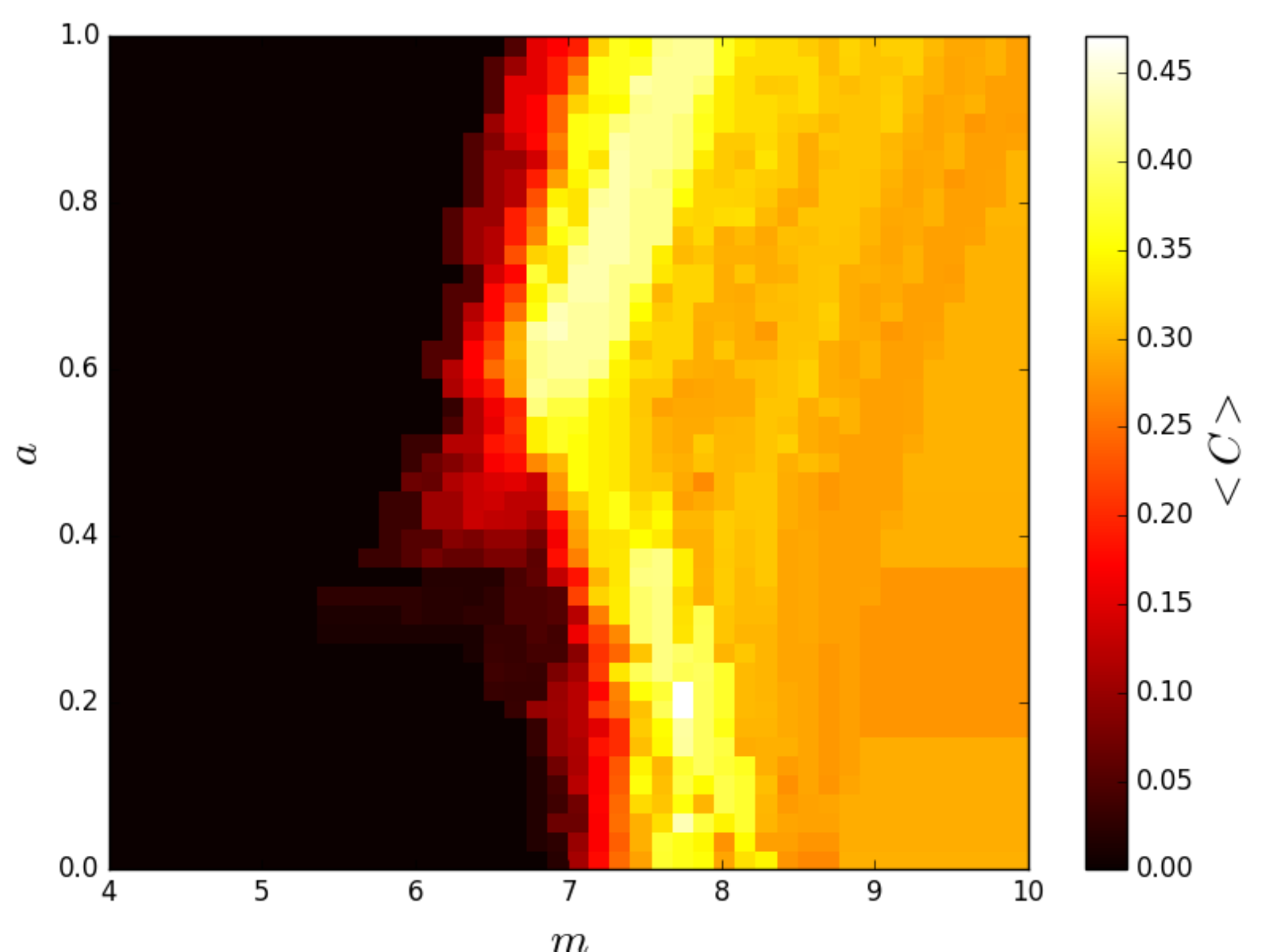} 
        \caption{Average clustering, $\langle C\rangle$} \label{fig:m1}
    \end{subfigure}
    \begin{subfigure}[h]{0.33\textwidth}
        \centering
        \includegraphics[width=\linewidth]{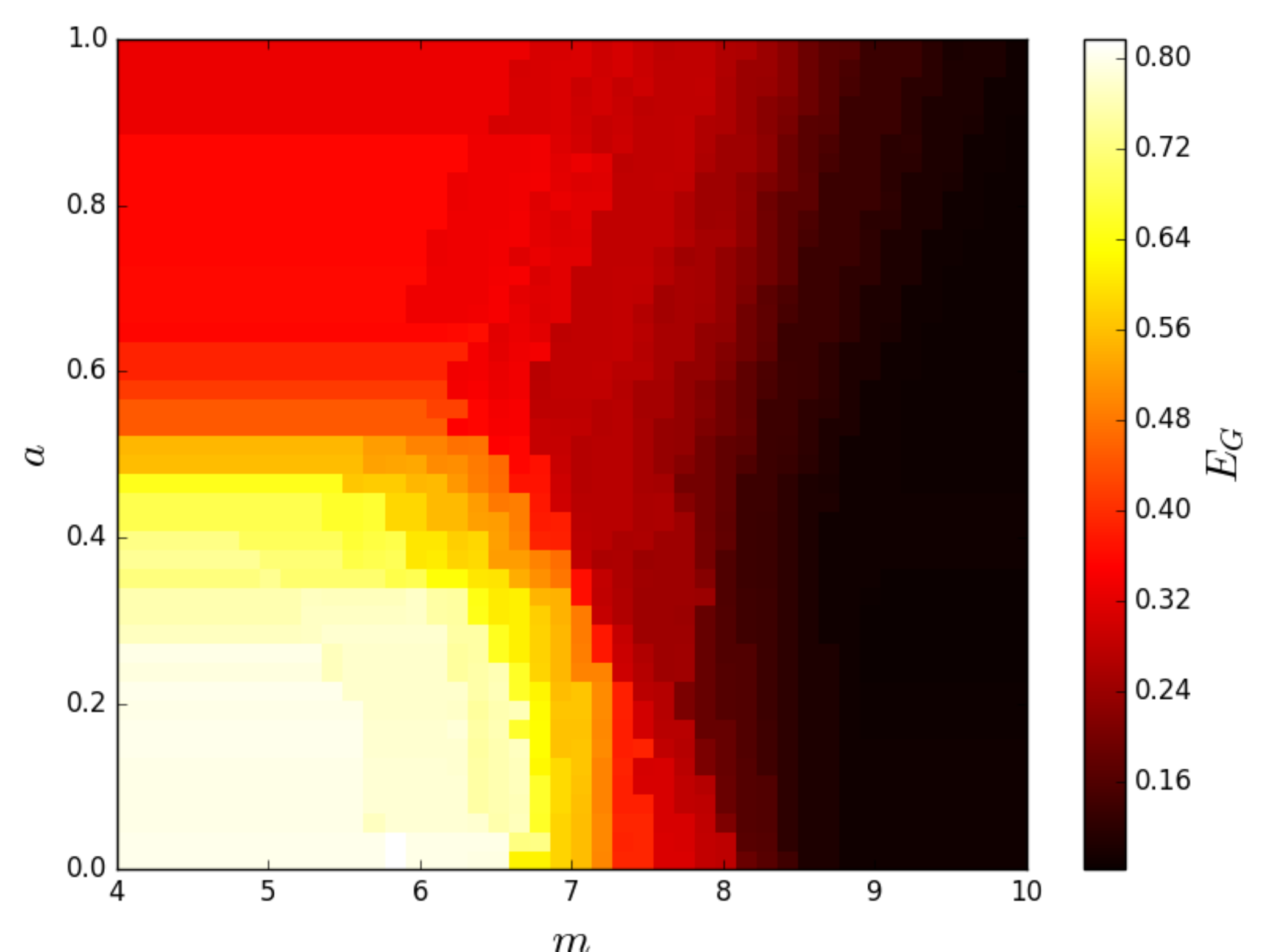} 
        \caption{Global efficiency, $\Eglob$} \label{fig:m2}
    \end{subfigure}
    \begin{subfigure}[h]{0.33\textwidth}
        \centering
        \includegraphics[width=\linewidth]{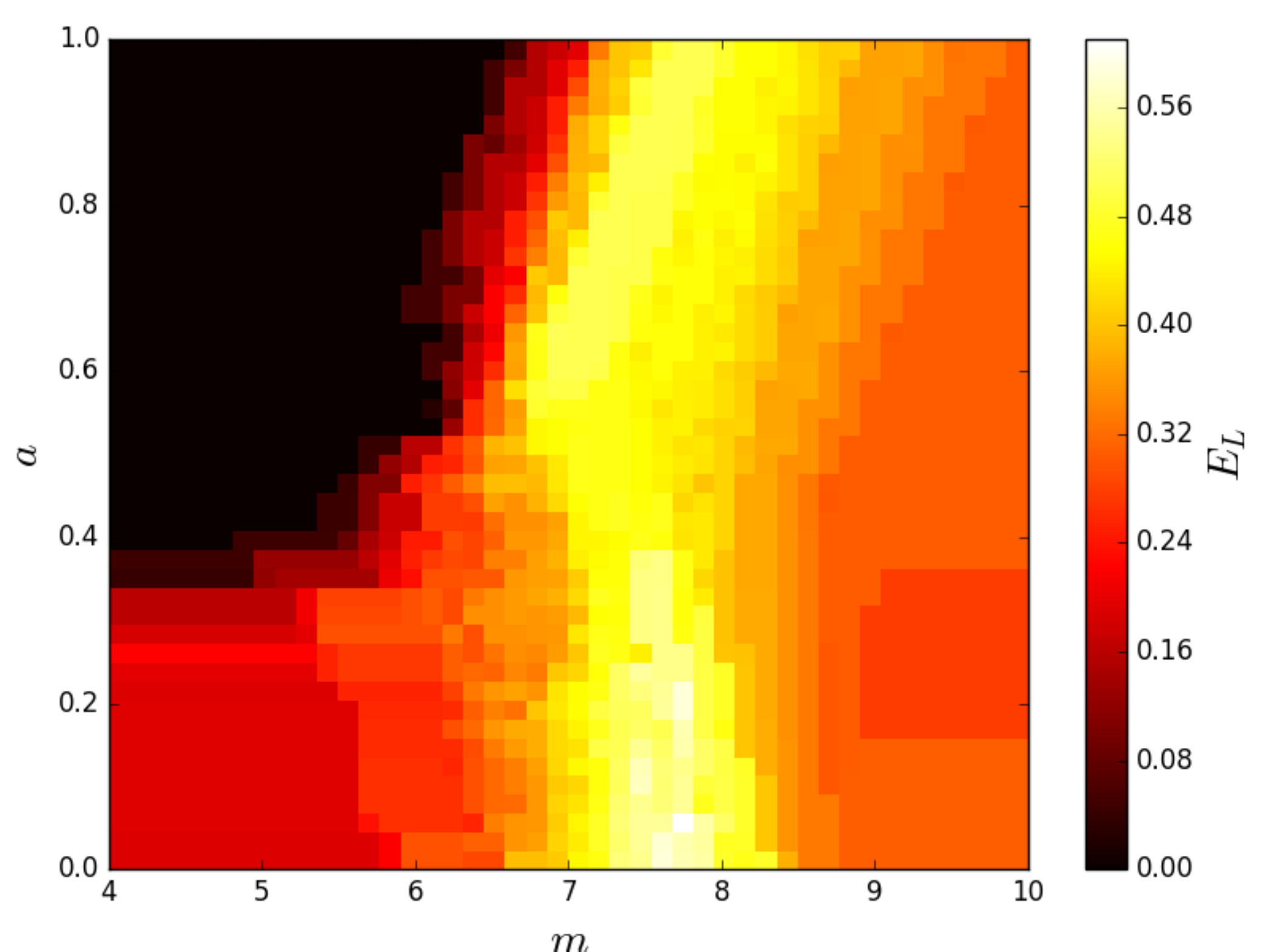} 
        \caption{Local efficiency, $\Eloc$}
        \label{fig:m3}
    \end{subfigure}

    \vspace{0.8cm}

    \begin{subfigure}[h]{0.33\textwidth}
        \centering
        \includegraphics[width=\textwidth]{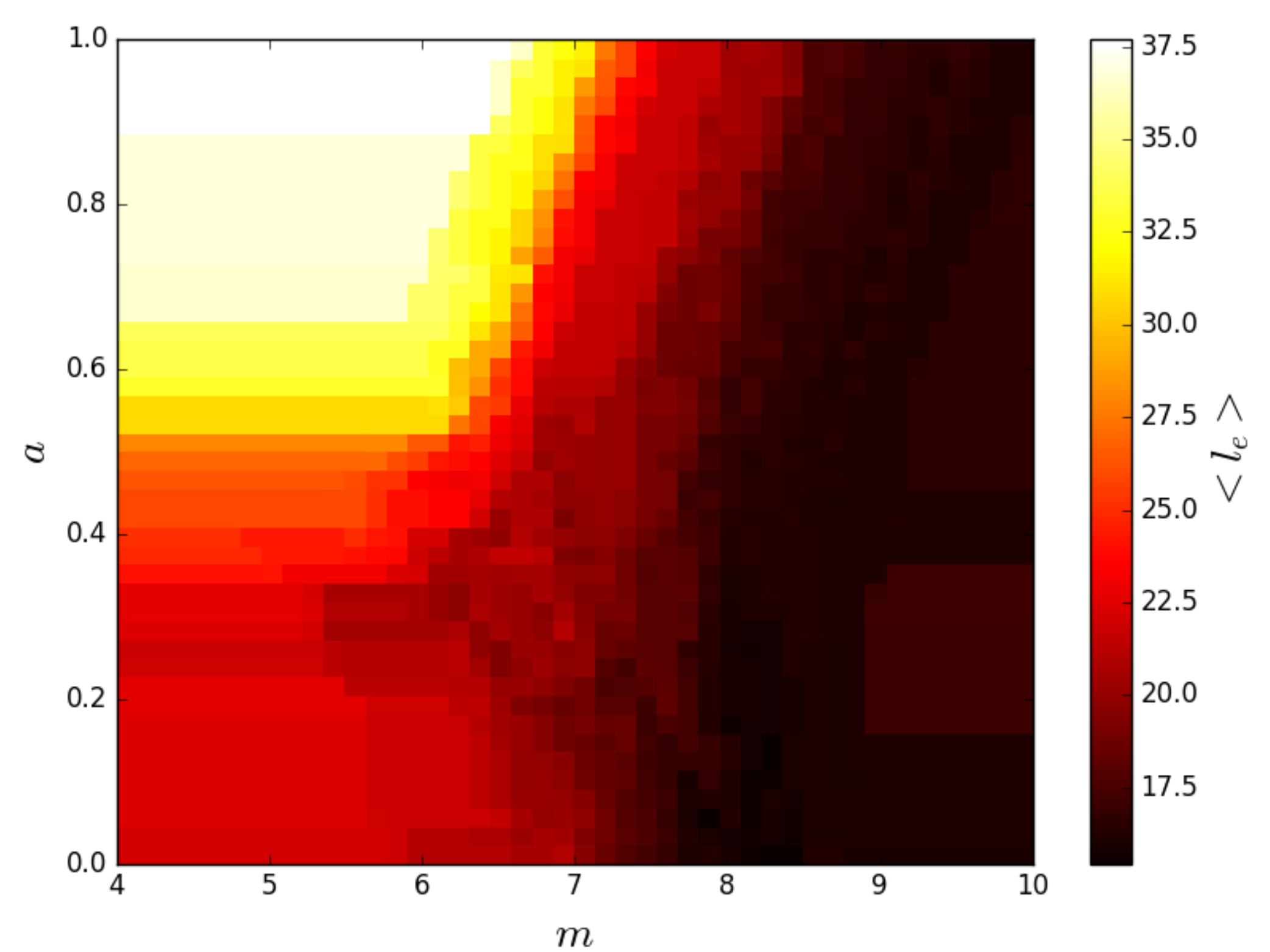}
    	\caption{Average link length, $\langle l_e\rangle$}
    	\label{fig:m4}
    \end{subfigure}
    \begin{subfigure}[h]{0.31\textwidth}
        \centering
        \includegraphics[width=\textwidth]{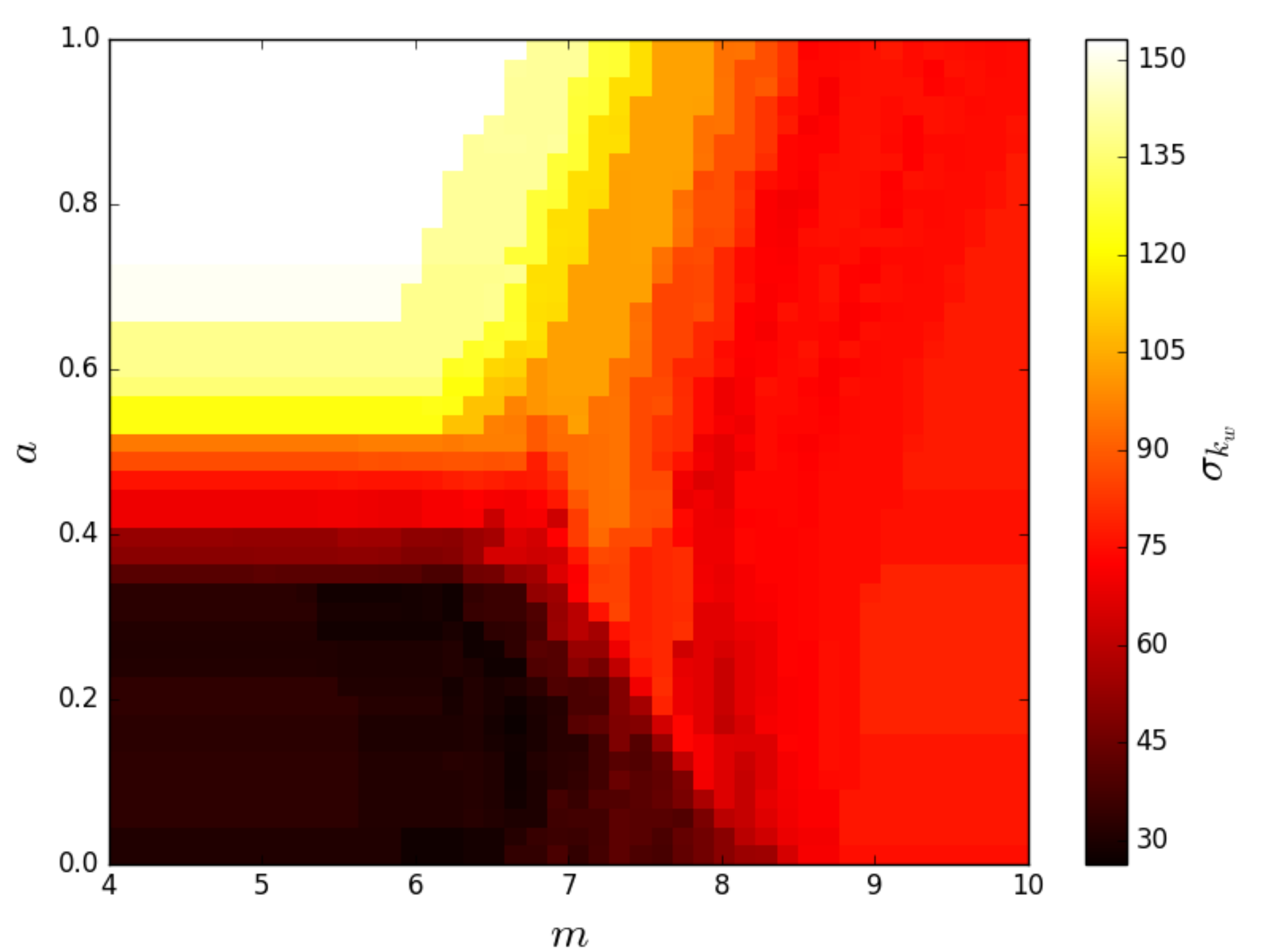}
    	\caption{Weighted degree standard\\ deviation, $\sigma_s$}
    	\label{fig:m5}
    \end{subfigure}
    \begin{subfigure}[h]{0.35\textwidth}
        \centering
        \includegraphics[width=\textwidth]{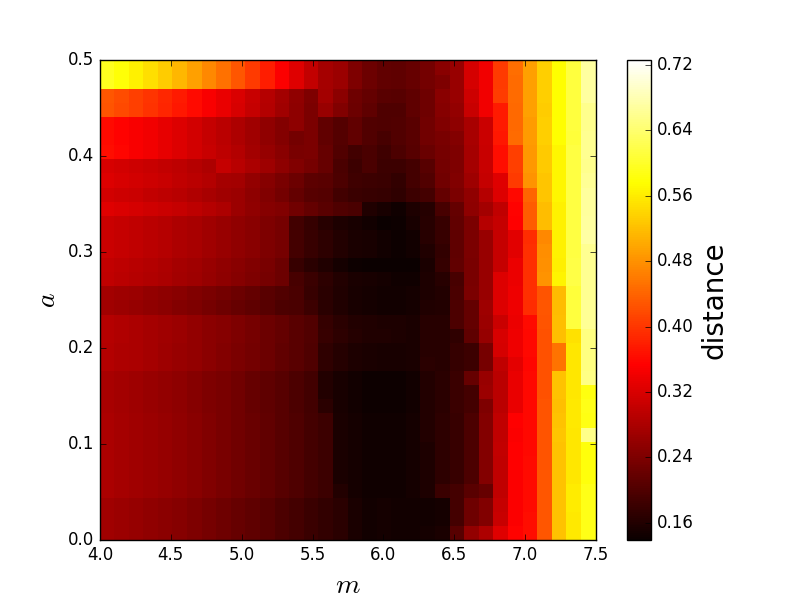}
    	\caption{SPL distance, $D_m$}
    	\label{fig:m6}
    \end{subfigure}

    \caption[Network properties in artificial networks.]
    {Network properties in artificial networks. \par \small For different values of $m$, $a$ artificial networks are created and their properties are computed.}
    \label{fig:metrics}
\end{figure}

The empirical network presents an almost vanishing clustering coefficient and a considerably high Global efficiency ($\Eglob^{emp} = 0.732$), just slightly lower than the EEM network (see Table \ref{tab:properties}). When comparing with the network topologies generated by the gEEM, we find that these two feature can be observed when $a \lessapprox 0.4$ and $0.55 \lessapprox m \lessapprox 0.65$ (see Figs.\,\ref{fig:m1} and \ref{fig:m2}). If we look at other metrics such as Local efficiency ($\Eloc^{emp} = 0.245$) the average link length ($\avg{l_e}^{emp}=21.7\,km)$, and the standard deviation of the node strength ($\sigma_s^{emp}=14.9\,km$), similar values can also be found in the same region of the parameter space.

Additionally, we computed a network distance between each synthetic topology and the empirical network based on the shortest path length between node pairs:

\begin{equation}
    D_m = \frac{2}{N(N-1)}\sum_{ij} 2\frac{|L_{ij}^{emp}-L_{ij}^{synt}|}{L_{ij}^{emp}+L_{ij}^{synt}},
\end{equation}

where $L_{ij}$ is the sum of the lengths of the links in the shortest path between $i$ and $j$ if both nodes belong to the same connected component, and $L_{ij}^{[d]} = e^{-m}L{CG}$ otherwise. For disconnected nodes in the empirical network, we assume the same value of $m$ that generated the synthetic counterpart we are comparing it to.

Once more, it is confirmed that the most similar topologies can be obtained by setting $m \in [0.6,0.65]$ while keeping the value of $a$ small, namely $a \lessapprox 0.35$.

For such values of $m$, we have $L_{ij}^{[d]} \in [93,153]\, (km)$, that is, larger than the average (geodesic) distance between node locations $\avg{d} = 83.5 \,(km)$, and also larger than the largest link in the empirical network $l_e^{max} = 85\, (km)$, but considerably smaller than the maximum distance between nodes $d_{max} = 269.5\, (km)$. 

In the system, for the selected value of merging parameter $\Gamma^*$, there are 12 nodes ($30.7\%$) whose average geodesic distance to other locations is in the range $[93,153]\, (km)$ or slightly larger.
 For these nodes, the underlying assumption of the gEEM that the overall preexisting connectivity allowed them to reach any other place in the system through a path of length $L_{ij}^{[d]}$ is, in this range of the parameter $m$, incorrect and, from the viewpoint of the infrastructure design, unfair.

In particular, the presence of isolates is a trait that can be observed in synthetic topologies only if $m>5$ and/or $a>0.35$.

Therefore, we can conclude that for the model to be able to reproduce this and other features of the empirical network, a non negligible share of unfairness towards the periphery of the system is required, either underestimating geographical distances ($m$), or explicitly favouring the most populated locations ($a$), or both. 
Taking into account the presence of alternative or preexisting TTIs in the same territory in a more realistic way ($m < 6$) does not produce synthetic networks that are close enough to the empirical one.

At a more local scale, the first node to get disconnected in the topologies generated by the gEEM is \textit{La Seu d'Urgell} (see Fig.\,\ref{fig:empirical}), which has the smallest merged population associated to it ($P=53154$) and the furthest distance to the nearest neighbor ($d=66.331\, km$). The isolates in the empirical networks have quite larger populations ($P(Palafrugell)=113773\ km$ and $P(Berga)= 84238$) and nearer neighbors (at $24.607\ km$ and $35.379\ km$, respectively). Although they both are far from the top ranking nodes, the features considered by the present version of the model do not make them plausible candidates to be the sole disconnected nodes of an otherwise connected topology. Possibly, the explanation for their specific condition could be found in the details of the historical process that shaped the TTI under study in its current state. For instance, until 1973, there existed a railway connection \textit{Berga}-\textit{Manresa} \cite{Perarnau-Llorens1981}, a link that the gEEM builds in most of the topologies that are similar to the empirical network.

\section{Conclusions}
We have presented a methodology to shed light on the mechanisms, power balances, and competing interests that shaped a given TTI into its current configuration. Relying on the analytical and theoretical toolbox provided by network science and adopting an inverse engineering approach, we propose a two-step procedure that allows us to address a broad variety of systems. 

The first step consists in mapping the infrastructure under study into a geographic network in a flexible way, adjusting the spatial scale of the representation to the specific situation of the system and its function by tuning a single parameter.

The second step enables us to investigate how different the considered system is from an ideal model of TTIs that has proven to be able to capture most of the relevant features of ancient proto-historical networks of pathways, that is, the EEM. The main characteristics of such a model are 
\begin{enumerate*}[label=(\roman*)]
\item the equitable treatment of the necessities and interests of each node-place in the system, regardless of its real power or importance;
\item the assumption that if in the network there is no path connecting two nodes, then it is impossible to reach one place from the other, that is, the considered network represent the only TTI existing in the territory.
\end{enumerate*}

We have devised a mechanistic network model, the gEEM, that using the EEM as a starting point, progressively and independently relaxes both its main assumptions by means of two parameters. 

Finally, we have shown how to effectively infer information about a real case-study applying the proposed methodology to an illustrative example, \ie the regional railway connections in Catalonia. Comparing the output of the gEEM for different values of the parameters to the empirical network of the Catalan regional train service, we evinced that network topologies similar to the empirical one can be generate if the most geographically central and demographically important places are slightly favoured when designing the TTI. These conclusions, which are in line with common knowledge but translate it into more precise quantitative terms, confirm the potential of simple mechanistic models as a powerful explanatory instrument for tackling complex systems.

\bibliographystyle{unsrt}
\bibliography{biblio2021}







\end{document}